# A NOVEL POINTING TECHNIQUE FOR THE ENHANCEMENT OF TROPOSPHERIC DELAY CALIBRATION SYSTEM PERFORMANCES

**David Bernacchia,[*] Riccardo Lasagni Manghi,[*†] Marco Zannoni, [*†] Paolo Tortora, [*†] Jose Villalvilla,[‡] Javier De Vicente,[‡] Paolo Cappuccio, [§] and Luciano Iess [**]**

The beam-crossing is a novel technique aimed at reducing residual tropospheric Doppler noise for micro-wave radiometer calibrations. In this work, we report the findings of the first test of this technique using ESA's Tropospheric Delay Calibration System (TDCS) at the complex in Malargue. The data consists in 14 tracking passes of the BepiColombo spacecraft collected between October 2023 and March 2024 during two separate test campaigns. We analyzed the performance of the beam-crossing technique and compared it with the nominal radiometer pointing through the analysis of the Doppler residuals extracted from the orbit determination process. Results show that the beam-crossing performed similarly to the standard pointing, with modest noise reductions and improved stability only at time scales between 100 s and 300 s. Key factors affecting the results include the antenna elevation and the boundary layer height, indicating the need to revisit initial test assumptions, which comprised a fixed boundary layer. Furthermore, comparing the beam-crossing test results with those obtained during the first two BepiColombo superior solar conjunction experiments highlights a potential application of this technique during periods of solar conjunction. However, technical challenges, adverse weather, and limited Ka-band transponder use, reduced the number of analyzed tracking passes. Future studies should therefore expand the dataset to consolidate the results. Furthermore new theoretical studies and test campaigns should elaborate on the selection process for the optimal crossing height.

## INTRODUCTION

The adoption of simultaneous usage of the X/X, X/Ka, and Ka/Ka frequency channels for radio tracking of spacecraft in deep space,  usually defined as a full multifrequency link, marked a substantial improvement in precision orbit determination.[16] In the context of signals calibration, the ability to eliminate the noise contributions coming from the dispersive media like the solar plasma and Earth's ionosphere,[1,2] shifted the attention towards troposphere-induced delay and delay rate, which constitute, with antenna mechanical noise, the most significant remaining noise contribution in Doppler measurements.

---

[*] Department of Industrial Engineering (DIN), University of Bologna, via Fontanelle 40, 47121, Forlì, Italy.
[†] Centro Interdipartimentale di Ricerca Industriale Aerospaziale (CIRI AERO) Alma Mater Studiorum – Università di Bologna, Via Baldassarre Carnaccini 12, 47121 Forlì, Italy.
[‡] European Space Agency, ESA-ESOC, Robert-Bosch-Straße 5, D-64293 Darmstadt, Germany
[§] European Space Agency, ESA-ESAC, Camino Bajo del Castillo s/n, 28692 Villafranca del Castillo, Madrid, Spain
[**] Sapienza University of Rome, via Eudossiana 18, Rome, Italy

Substantial progress in mitigating tropospheric noise has been made by replacing standard GNSS-based calibrations with measurements performed by water vapor radiometers installed close to the ground tracking antenna, retrieving the path delay induced by the water vapor along the antenna's line of sight [3].

In the case of the European Space Agency, the prototype of a new water vapor radiometer, the Tropospheric Delay Calibration System (TDCS), has been installed close to the deep space ground station DSA-3 in Malargue, Argentina.[4] This new calibration system has been tested in the context of the ESA missions Gaia and BepiColombo, showing improved tropospheric noise removal with respect to GNSS calibrations.[5,6] Despite its promising results, the TDCS exhibits limitations that hinder its noise cancellation performance. In particular, these limitations are mainly due to the geometric configuration of the radiometer, being separated from the ground antenna by about 30 m. This spatial offset causes the instrument to observe a different portion of the atmosphere with respect to the one observed by the antenna. This discrepancy is also increased by the difference between the shapes of the DSA and the radiometer's beams.[7]

To assess the error introduced by this geometric condition, an analysis of the data collected during the first two Solar Conjunction Experiments (SCE) was conducted as part of the Mercury Orbiter Radioscience Experiment (MORE), one of the science investigations of the mission Bepi-Colombo.[8,9,10] The analysis revealed the presence of uncalibrated (or residual) tropospheric Doppler noise extracted from the orbit determination (OD) process.[11] The uncalibrated noise was particularly evident during passes characterized by superior solar conjunction, where the Sun avoidance mode was activated. This feature of the radiometer is enabled when the Sun-Earth-Probe (SEP) angle is small ($< 3°$) to prevent direct sunlight from interfering with the instrument's measurements. The mode operates by applying a deliberate pointing offset from the nominal configuration of the TDCS, which nominally points in the same direction of the main antenna beam. This offset is designed to minimize the angular deviation, regardless of the direction in which it is performed.

In this work, we illustrate the outcomes of the first-ever testing of the beam-crossing method, a novel radiometer pointing technique aimed at reducing the error due to the misalignment between the TDCS and the ground station antenna. This technique, described theoretically but never tested on the field, requires the radiometer to establish an angular pointing offset to cross the antenna line of sight at a desired altitude[7] rather than maintaining the standard parallel configuration. This particular altitude should correspond to the one that minimizes the discrepancy between the observed tropospheric water vapor contents. In this way, the error due to the spatial separation between the two beams can be reduced, resulting in a decreased tropospheric noise level and improving the accuracy of the orbit determination process.

The main difference between the beam-crossing configuration and the sun avoidance mode is that in the former the pointing offset is directed towards a specific desired direction rather than being oriented to minimize the angular deviation.

Figure 1 illustrates a schematic representation of the beam-crossing geometry compared to the nominal parallel pointing. As represented in the picture on the left, the main idea behind this technique is to make the radiometer observe the volume of the atmosphere followed by the antenna where one expects to find the most significant contribution to tropospheric noise.

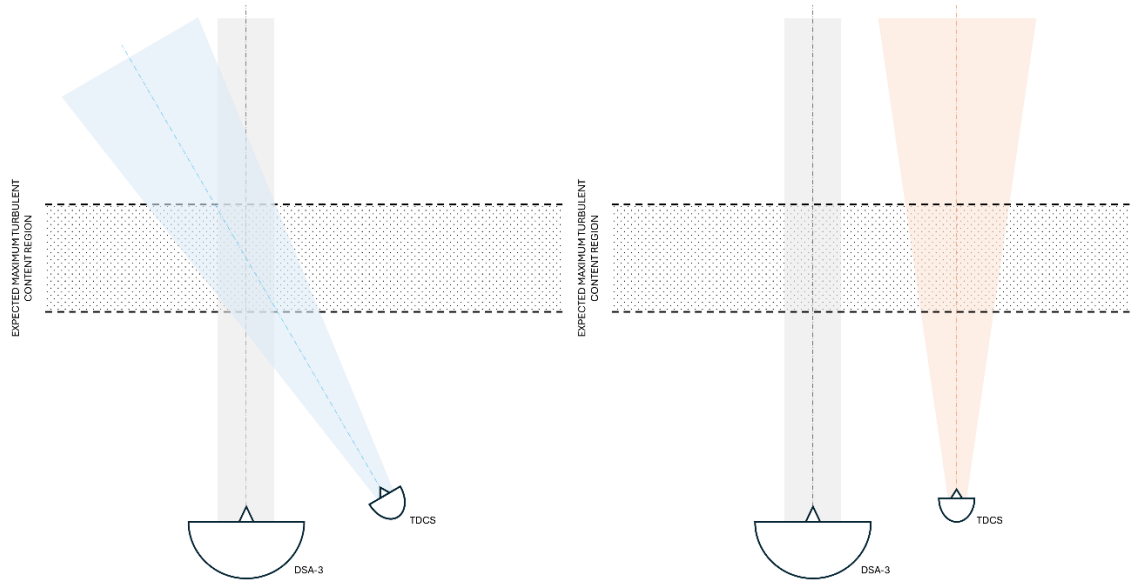

**Figure 1 Conceptual scheme of the beam-crossing (left) compared to the nominal pointing (right).**

## TEST SETUP

The beam-crossing pointing technique was tested in the frame of the BepiColombo mission during two different testing campaigns: the first one performed between October and November 2023, comprising a total of 12 tracking passes, and the second one between February and March 2024, consisting of 3 tracking passes. For all 15 passes, the spacecraft tracking had been operated from the ESA DSA-3 ground station.

The main limitation of the experiment was the inability to directly compare the standard pointing with the beam-crossing due to the availability of only one TDCS at the ground station. To address this, the water vapor radiometer was used in a specific configuration, in which the standard pointing and the beam-crossing were alternated during the same tracking session, trying to minimize the variability of the external conditions by using a fast switching rate. During the first campaign, the switching time between the two pointing strategies was set at 1 hour, leading to piecewise continuous tracking sessions of the same duration. This specific timing was initially chosen to provide a sufficiently large time window for characterizing the Doppler stability, usually evaluated through the Allan deviation, at long stability intervals (i.e., 1000 seconds). The setup was then modified during the passes of the second campaign, where the switch time was reduced to 15 minutes. The main reasons behind this decision were the need to increase the number of segments with the same pointing to have a more statistically reliable dataset, and to decrease the effects of meteorological variability. The trade-off for reducing the time window of a single segment has been renouncing to the possibility of having a reliable estimate of the Allan deviation for high integration time values. Consequently, we established a maximum integration time of $\tau = 300$ seconds as a reference point for the Allan deviation analysis.

Concerning the definition of the crossing point, as stated before, the ideal case scenario would be to have the crossing occurring in correspondence with the zone where we expect to find the most turbulent content, hence the region introducing the most tropospheric noise into the radiometric observables. Different studies showed that the boundary layer plays a crucial role in atmospheric turbulence, acting as the primary region where turbulent processes occur,[12,13] so having the crossing in correspondence with this particular portion of the atmosphere was a first design requirement.

However, the height of the boundary layer can range from tens to thousands of meters, and it is characterized by high variability, both diurnal and seasonal, being strictly related to meteo conditions. Moreover, identifying which portion of the boundary layer is the one introducing the most significant tropospheric noise is not a straightforward task. For this reason, the work described by Graziani[7] was used as a reference to determine the point for the crossing of the beams. The analysis, conducted in the framework of the AWARDS study commissioned by ESA,[14] involved different simulations of a beam-crossing scenario in which it was identified that the condition to reach the highest tropospheric noise removal was to have the crossing occurring at the 70% of the boundary layer height so, assuming a boundary layer height of 1 km the ideal crossing height would be 700 meters. Despite the simplified setup used for the simulations, which differed significantly from real-world conditions, we used this value as a preliminary reference for our analysis, also in the optics of validating the simulations' results. Consequently, the 700-meter height was used as the crossing reference point for both campaign #1 and campaign #2. The validity of this assumption will be discussed in detail in the results section.

Table 1 summarizes the principal setup parameters used in the two testing campaigns.

**Table 1 Setup parameters for beam-crossing testing.**

|  | Switching interval [s] | Crossing height [m] |
|---|---|---|
| Campaign #1 | 3600 | 700 |
| Campaign #2 | 900 | 700 |

## ORBIT DETERMINATION ANALYSIS

The core of the analysis consists in a comparison of the two pointing strategies, obtained by performing an orbit determination (OD) process with the Doppler and Range data acquired during the different tracking passes. Since a direct comparison of the two strategies is impossible, the orbit determination process was performed twice, first applying the tropospheric calibrations obtained by the TDCS and second applying tropospheric calibration retrieved from GNSS data. In this way, the percentage of noise reduction from GNSS calibration to TDCS could be used as a further evaluation factor. The OD setup was fixed between the two analyses to keep the tropospheric calibration as the only variable and avoid additional error contributions.

### Data selection and processing

During the October-November 2023 test session, the spacecraft's Ka-Transponder (KaT) was activated, establishing the full multifrequency link. Currently, BepiColombo is in a critical phase of the mission as it is approaching its orbital insertion around Mercury. Additionally, the scientific plan for the cruise phase included a Solar Conjunction Experiment (SCE) in December 2023, making the spacecraft's operational schedule tight. For this reason, only 12 passes with the full multifrequency link enabled were allocated for the beam-crossing tests out of which one pass was removed due to a technical malfunction of the TDCS, resulting in 11 passes being included in the final analysis.

The schedule issue described above also prevented operating the KaT during the second test campaign, which exploited three already scheduled TT&C tracking passes. The standard Bepi-Colombo TT&C tracking passes from Malargue are performed with the dual link configuration comprising the X-up/X-down and X-up/Ka-down, which allows for a complete removal of the downlink dispersive noise only.[1,2] Table 2 offers a summary of the 14 passes that were comprised in the final analysis, including information about the available frequency bands and about the length of the tracking sessions. The full multifrequency link is indicated as "Triple link" while the X/X and X/Ka configuration is referred as "Dual link".

**Table 2 Tracking passes summary.**

| Year and Day Of Year | Date | Frequency link | Begin of data (UTC) | End of data (UTC) |
|---|---|---|---|---|
| 2023298 | 25 Oct 2023 | Triple | 10:48:19 | 13:05:43 |
| 2023300 | 27 Oct 2023 | Triple | 10:48:03 | 15:10:59 |
| 2023305 | 01 Nov 2023 | Triple | 12:09:59 | 15:13:08 |
| 2023307 | 03 Nov 2023 | Triple | 10:48:11 | 15:14:51 |
| 2023308 | 04 Nov 2023 | Triple | 10:48:17 | 14:39:50 |
| 2023312 | 08 Nov 2023 | Triple | 10:48:40 | 15:16:56 |
| 2023313 | 09 Nov 2023 | Triple | 10:48:47 | 14:41:51 |
| 2023314 | 10 Nov 2023 | Triple | 10:48:55 | 15:26:13 |
| 2023315 | 11 Nov 2023 | Triple | 10:49:03 | 13:29:59 |
| 2023319 | 15 Nov 2023 | Triple | 10:49:44 | 13:24:11 |
| 2023321 | 17 Nov 2023 | Triple | 10:50:12 | 15:09:38 |
| 2024053 | 22 Feb 2024 | Dual | 10:55:58 | 18:19:11 |
| 2024074 | 14 Mar 2024 | Dual | 10:26:43 | 16:45:08 |
| 2024081 | 21 Mar 2024 | Dual | 10:27:16 | 16:41:25 |

The Doppler and range data registered by the ESA DSA-3 station during the different passes, delivered in TTCP format,[15] had been processed to remove the delays introduced by the electronic and optical systems of the ground station. Information about these delays are usually retrieved during dedicated calibration sessions before and after the pass and delivered in TTCP files. A second calibration step included the removal of the delays introduced by the spacecraft electronic system, mainly the Deep Space Transponder (DST) and the KaT (only for the test campaign #1). In particular, the delays of the Ka-Transponder are measured thanks to the self-calibration capability of the instrument and the outcome data can be retrieved from the spacecraft's telemetry. In the case of the DST, the available informations are collected in technical reports summarizing the results of on-ground calibration campaigns conducted before launch. Finally, the last processing steps included the calibration of media effects (described in a following dedicated section), the removal of all the data at elevations below 15° and the elimination of eventual outliers through a visual inspection of the residuals resulted from the OD analysis.

**Media calibration**

In deep space tracking, the main noise contributions affecting radiometric observables are typically introduced by dispersive media such as the solar plasma and the Earth's ionosphere. The radio system embedded in the BepiColombo probe is a potent tool for the removal of these effects since the contemporary usage of two different transponders, the DST and KaT, which enable the already mentioned full multifrequency link, allows for the mathematical cancellation of the dispersive noise from the observables.[1,16,17] It is possible to demonstrate that through the linear combination of the observables coming from the different frequency channels, the non-dispersive contribution of the measurements can be isolated, removing the dispersive noise part. Considering Doppler and range observables as the combination of a non-dispersive component $y_{nd}$ and two components related to the uplink and downlink carrier frequencies $f_\uparrow$ and $f_\downarrow$, we can write:

$$y = y_{nd} + \frac{P_\uparrow}{f_\uparrow^2} + \frac{P_\downarrow}{f_\downarrow^2} \tag{1}$$

The terms $P_\uparrow$ and $P_\downarrow$ in Equation (1) are coefficients proportional to the media's total electron content along the signal's path (electrons/m²) in the case of range measurements and proportional to its time derivative in the case of Doppler ones. Introducing the turnaround ratio $\alpha = f_\downarrow / f_\uparrow$ and defining $y_u = P_\uparrow / f_\uparrow^2$ and $y_d = P_\downarrow / f_\uparrow^2$, Equation (1) can be rewritten as:

$$y = y_{nd} + y_u + \frac{y_d}{\alpha^2} \tag{2}$$

Writing this expression for the measurements obtained through the three different frequency channels, namely $y_{xx}$, $y_{xk}$ and $y_{kk}$, we can solve the system of the 3 equations in the 3 unknowns $P_\uparrow$, $P_\downarrow$ and $y_{nd}$ where the latter is exactly the plasma-free component of the observables. The final expression for this non-dispersive part is:

$$y_{nd} = \left(\frac{1}{\beta^2 - 1} \frac{\alpha_{xx}^2}{\alpha_{kk}^2} \frac{\alpha_{xk}^2 - \alpha_{kk}^2}{\alpha_{xx}^2 - \alpha_{xk}^2}\right) y_{xx} + \left(\frac{1}{\beta^2 - 1} \frac{\alpha_{xk}^2}{\alpha_{kk}^2} \frac{\alpha_{kk}^2 - \alpha_{xx}^2}{\alpha_{xx}^2 - \alpha_{xk}^2}\right) y_{xk} + \left(\frac{\beta^2}{\beta^2 - 1}\right) y_{kk} \tag{3}$$

In the case of the dual link, the absence of the Ka/Ka-band does not allow this mathematical removal of the dispersive noise content. However, being the X-band uplink used both for the X/X and X/Ka channels, the uplink term $y_u$ is the same for both the measurements $y_{xx}$ and $y_{xk}$. It is trivial to demonstrate that, thanks to this equality in the component of the plasma uplink for the two observables, it is possible to mathematically isolate the contribution of the plasma in the downlink, given by the following equation:

$$y_d = \frac{y_{xx} - y_{xk}}{\alpha_{xx}^{-2} - \alpha_{xk}^{-2}} \tag{4}$$

Since only the downlink plasma contribution can be retrieved mathematically, different methods have been developed to estimate the uplink contribution, such as the "thin screen" method. The thin screen is a simplified model that is particularly effective in estimating the uplink plasma content at low values of the SEP angle, while it becomes more inaccurate as the elongation angle increases.[18] In the case of the beam-crossing test, as depicted in Figure 2, during the time window of the second campaign, the SEP was particularly low in correspondence with the 2024053 pass, while for the 2024074 and 2024081 passes, it grew up to values around 35°. In this particular condition, in the case of the 2024053 pass the thin screen offered a better calibration performance while for the other two passes accounting for the only downlink plasma contribution resulted in less noisy Doppler residuals. This calibration scheme was the one used for the final analysis.

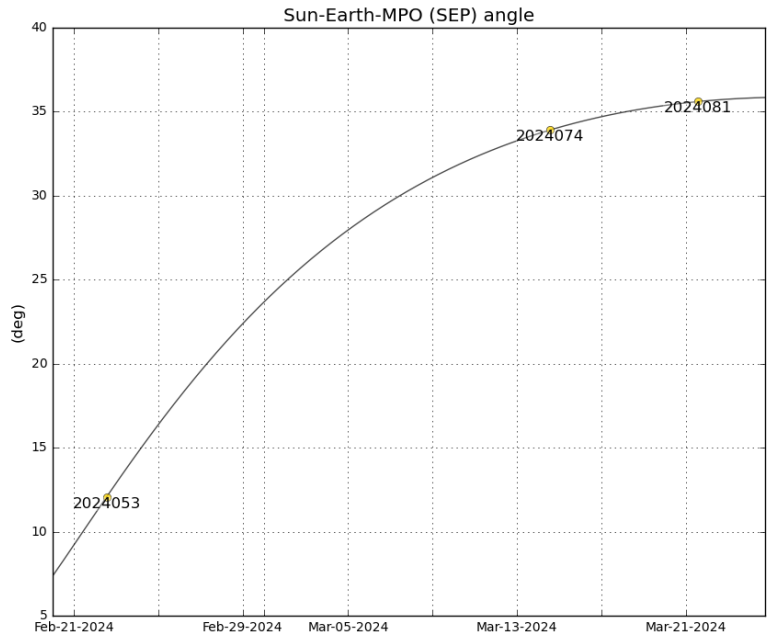

**Figure 2 SEP angle during the beam-crossing testing campaign #2.**

Concerning the tropospheric noise, the two different types of calibration, GNSS and TDCS, have been used in separate OD analyses. The standard GNSS calibrations were generated by the ESOC Navigation Support Office[19] and provided in CSP format. The TDCS calibrations are obtained by processing the sky brightness temperature measurements collected by the radiometer and stored in CSP format. Further details about the two different calibration types can be found in Lasagni Manghi *et al.* 2023.[6]

**Dynamical model and filter setup**

The orbit determination was conducted using the Mission Analysis, Operations, and Navigation Toolkit Environment (MONTE) software developed by NASA's Jet Propulsion Laboratory.[20] This software employs a weighted least-squares batch filter to minimize residuals, defined as the difference between observed and simulated measurements, by adjusting the values of the solved-for parameters.

The dynamical model used to compute the simulated measurements comprises the point-mass gravitational perturbations of the Sun, the solar system's planets, and their satellites. In the case of Mercury and the Sun, the model was expanded with higher-order spherical harmonics. The state vectors of the different bodies and their gravitational coefficients were retrieved from JPL's DE440 planetary ephemerides.[21]

Regarding non-gravitational accelerations, the main contribution is given by the solar radiation pressure (SRP), which was included in the model. This was implemented using a standard flat plates model and assuming a polyhedral shape for the main spacecraft components. Surface thermo-optical properties were modeled to vary linearly from the launch to the end of the mission. The spacecraft's attitude information used to compute the SRP was extracted from the BepiColombo operational SPICE kernels provided by the ESA SPICE Service.[22] Additional, smaller, non-gravitational accelerations such as the Thermal Recoil Pressure (TRP) or the Solar Wind were neglected since their effect is negligible on the time scale of the duration of a single tracking pass.[26]

The solved-for parameters estimated in the OD process were the spacecraft's position and velocity vectors, the SRP scale factor, the ground station range bias, and the spacecraft's center of mass relative to the body frame. Further details about the parameters' *a priori* values and uncertainty are given in Reference 6, where the same orbit determination setup was used to analyze the data acquired during the first two solar conjunction experiments performed by BepiColombo.

## METHODS

The effectiveness of the beam-crossing (BC) technique was assessed comparing its performance to that of the standard pointing (SP). This comparison was made by examining the noise characteristics of the Doppler residuals obtained from the orbit determination process. To facilitate the comparison, the residuals were split into different segments characterized by the same pointing strategy.

Initially, the characteristics of the different segments were analyzed individually within the context of a single pass. Subsequently, a comprehensive analysis was conducted by aggregating data from all 14 passes. Furthermore, the noise characteristics of the residuals were correlated with different variables related to the pointing geometry, including the antenna's azimuth, its elevation, and to the local meteorological conditions at the station. The meteo data has been retrieved from different sources including the weather station embedded in the TDCS,[5,6] the WEST weather station collocated at the Malargue site, and the ERA5 global reanalysis dataset provided by the European Center for Medium-range Weather Forecast (ECMWF).[23,24]

Finally, residuals' statistics obtained with the beam-crossing technique were compared with those obtained during BepiColombo's SCE1 and SCE2 for the tracking passes when the *Sun avoidance* mode was activated. Similarly to the beam-crossing, the *Sun avoidance* mode foresees an intentional off-pointing of the TDCS with respect to the ground station antenna with the goal of avoiding possible intrusions of the Sun within the TDCS beamwidth. Having different goals, the two techniques may result in different pointing directions and, consequently, calibration performances.

## RESULTS

Figure 3 shows, as example, the Doppler residuals at 60s for the tracking pass performed on November 3[rd], 2023 (DOY 307). This pass belonged to the first testing campaign and had a switch time of 1-hour. Following the color code used in Figure 1, the red dots represent the residuals obtained using the standard pointing while the blue dots are the residuals relative to the beam-crossing strategy.

It should be noted that the initial and final intervals have a shorter duration then the others. This is due to a discrepancy between the scheduled switching times set by the radiometer and the actual tracking interval commanded by the ground station antenna. Start or end of tracking before the conclusion of the full 1-hour batch may result in a few segments being truncated. From Figure 3 we can notice that the first two complete segments are the noisiest of the entire pass, likely as a result of the higher integrated water vapor content at low elevations. However, a difference in the overall performances of the two strategies is difficult to determine from directly observing the data.

Figure 4 shows the Allan deviation (ADEV) of the Doppler residuals at 1s relative to the different segments of the tracking pass in Figure 3. We can see that the various segments show different ADEV values for time scales above roughly 10 seconds while still satisfying the MORE requirements in terms of stability.

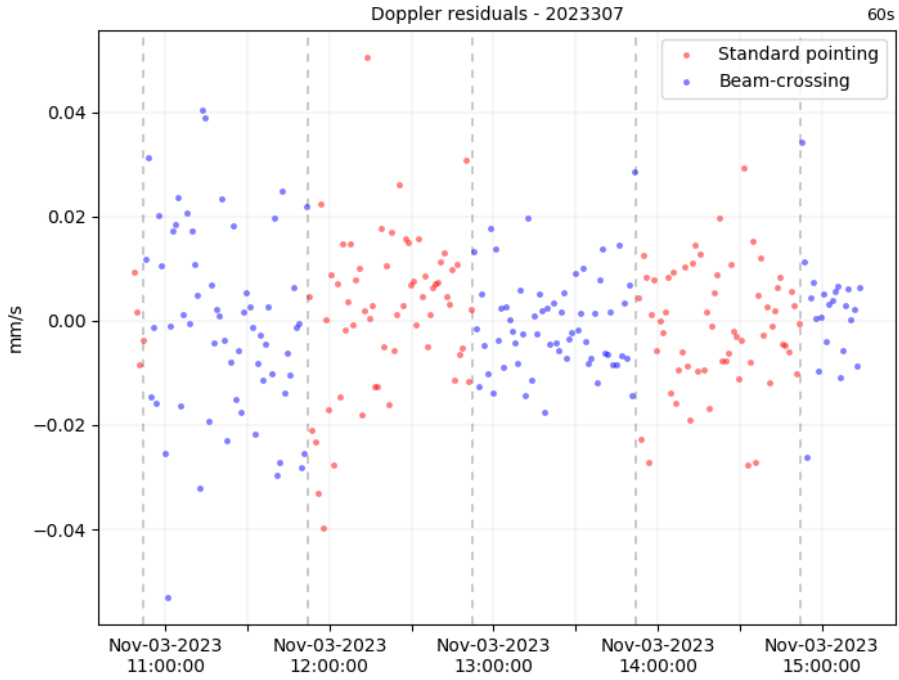

**Figure 3** Doppler residuals at 60s obtained during tracking pass 2023307. Red dots: standard pointing; blue dots: beam-crossing. Vertical lines: switching times between pointing strategies.

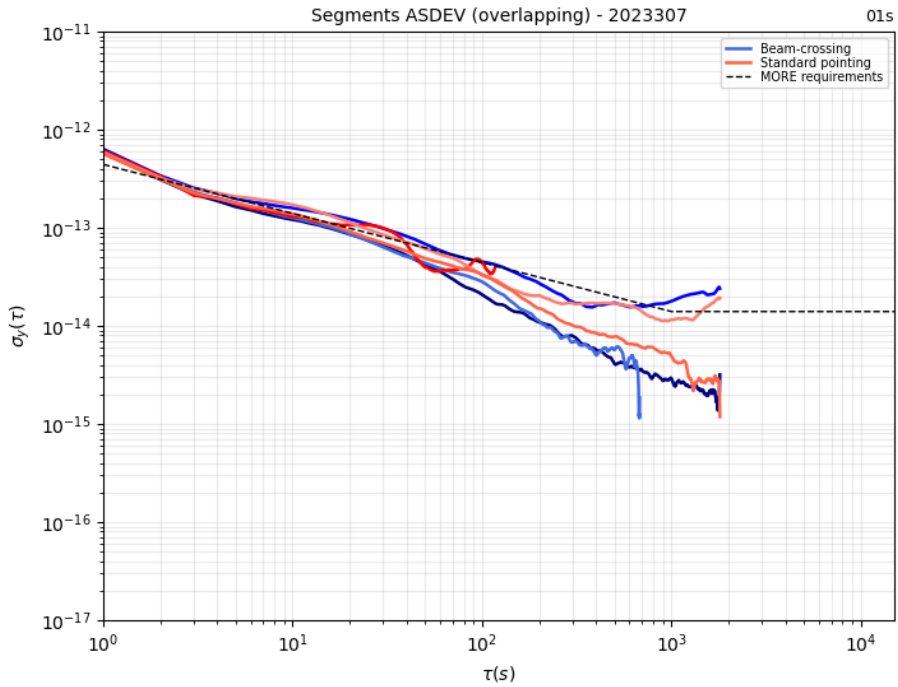

**Figure 4** Allan deviation of the Doppler residuals at 1s count time for pass 2023307. Red lines: ADEV of the standard pointing segments; blue lines: ADEV of the beam-crossing segments; black dashed line: MORE stability requirement.

The same visual examination was repeated for all passes (not reported here for brevity) and the overall statistics of different segments belonging to the same pointing strategy were collected and compared globally.

To facilitate the comparison between segments collected in the two campaigns, characterized by different switching rates, the 1-hour segments of the campaign #1 were split in 15-minutes intervals. The same procedure was repeated also for the residuals calibrated with GNSS-based tropospheric calibration. In this way, the segments were catalogued according to the time window in which they occurred, namely beam-crossing time window and standard pointing time window.

The outcome of this global RMS analysis is presented in Figure 5, where the dark bars represent the RMS levels of the TDCS-calibrated residuals while the shaded bars are relative to the GNSS-calibrated data. It can be seen that the noise levels of the TDCS-calibrated measurements is comparable for the two cases, with the beam-crossing and standard pointing having nearly identical RMS values. Conversely, the corresponding GNSS-calibrated data is characterized by slightly higher noise level on the beam-crossing intervals, resulting in a noise reduction of 34%, compared to the 32% reduction achieved by standard pointing. The numerical values are reported in Table 3.

**Table 3 Global RMS values.**

|  | TDCS calibration [mm/s] | GNSS calibration [mm/s] | Noise reduction percentage |
|---|---|---|---|
| Beam-crossing | 0.01909 | 0.02893 | 34.01 % |
| Standard pointing | 0.01930 | 0.02842 | 32.09 % |

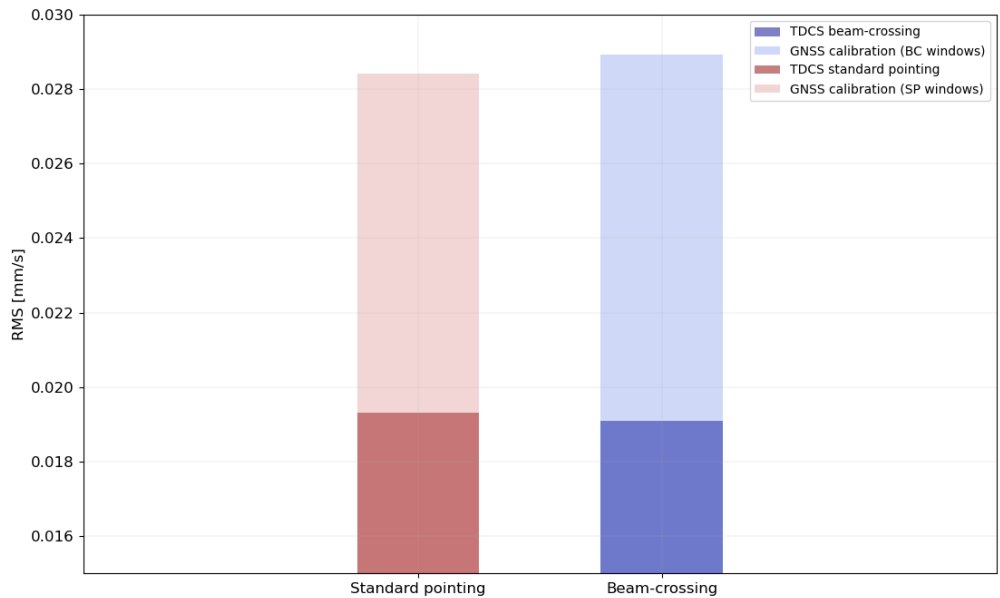

**Figure 5 Global RMS values for the two pointing strategies compared with GNSS-calibrated residuals. Red: intervals related to standard pointing; blue: intervals related to beam crossing.**

A second step in the global analysis comprised the examination of the Allan deviation. Similarly to the RMS case described before, we computed the average value of the Allan deviation for the ensemble of tracking segments computed with a given pointing strategy. In particular, we used as reference characteristic integration times of 1s, 20s, 60s, 100s, and 300s. The limiting factor for the maximum value of integration time has been the switching rate of 15-minutes between the two strategies. Figure 6 shows the average ADEV values for the two strategies as a function of the stability interval along with the maximum and the minimum values registered among the different segments.

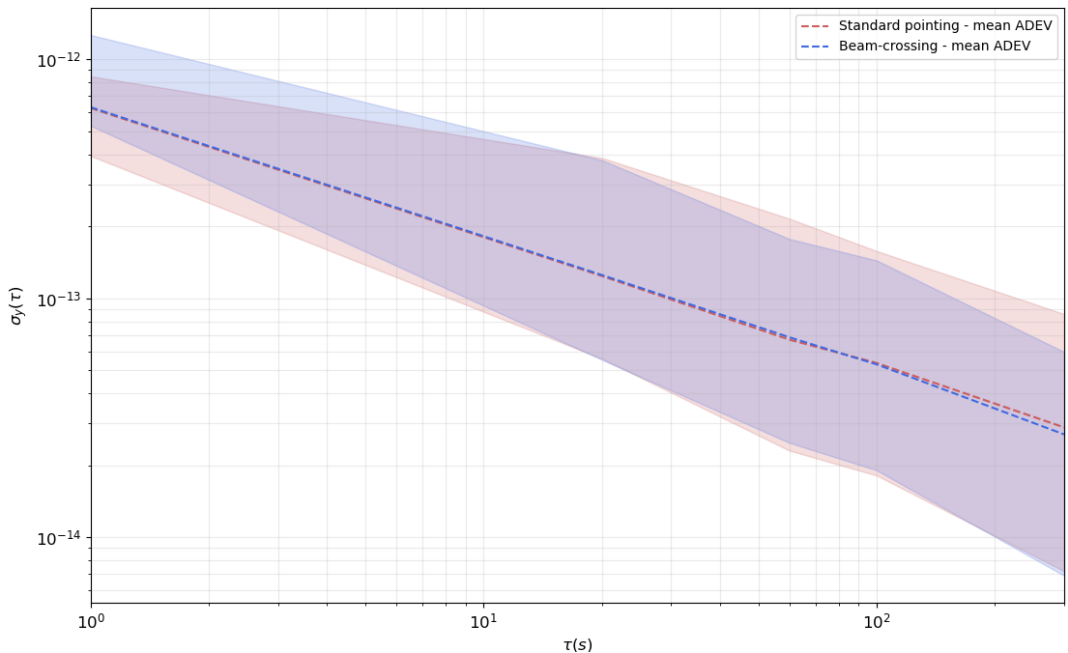

**Figure 6 Allan deviation statistics over the ensemble of tracking segments. Dashed lines: mean ADEV values; shaded areas: maximum and minimum values. Red: standard pointing; blue: beam crossing.**

In general, the two strategies provided similar performances, with mean value curves almost overlapping with the exception of stability intervals between 100 s and 300 s, where we observe slightly lower values with the beam-crossing. The main difference is observed in the extreme values where we observe that the beam-crossing performs worse than the standard pointing at low stability intervals and better at higher stability intervals (> 20 s).

Once analyzed the general noise characteristics of the residuals, we performed a correlation analysis to determine the influence of different variables on the performances of the two strategies. Different parameters were taken into account, including elevation of the antenna, wind speed, wind direction, rain rate, and liquid water path. Analogously to the residuals, time series of these parameters were split in 15-minutes segments corresponding to the same intervals covered by the two pointing strategies.

The different variables were studied in relation with the absolute RMS of the TDCS-calibrated residuals and with the RMS ratio between TDCS-calibrated measurements and GNSS-calibrated measurements, namely:

$$RMS_{RATIO} = \frac{RMS_{TDCS}}{RMS_{GNSS}} \tag{5}$$

The most relevant results of this analysis are those related to the correlation of the RMS and RMS ratio with the elevation of the ground station antenna, which are shown in Figure 7 and Figure

8. In these figures, values marked with T represent tracking segments during which the full multifrequency link was active (campaign #1) while those marked with D indicate segments during which only a dual link was used (campaign #2).

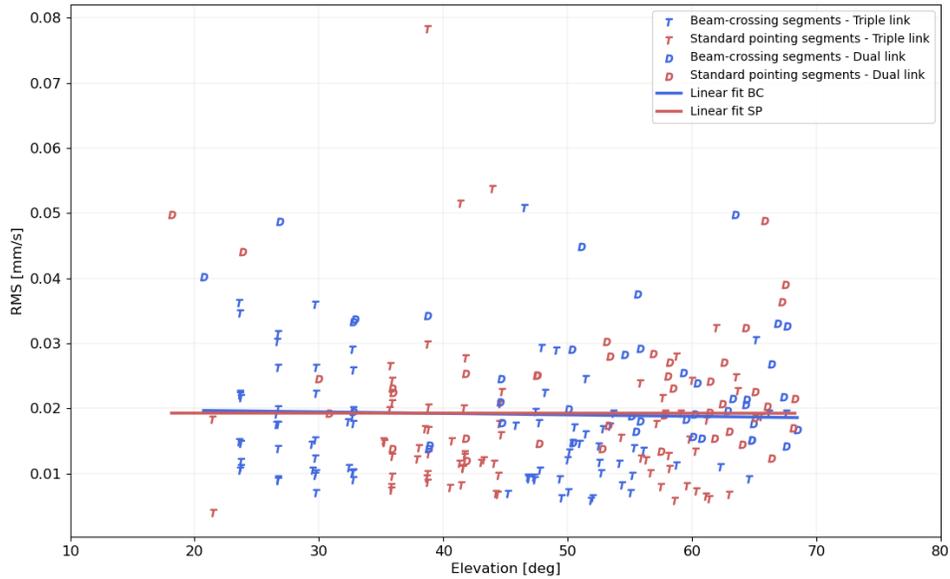

**Figure 7 RMS of TDCS-calibrated residuals as function of the ground station antenna elevation.**

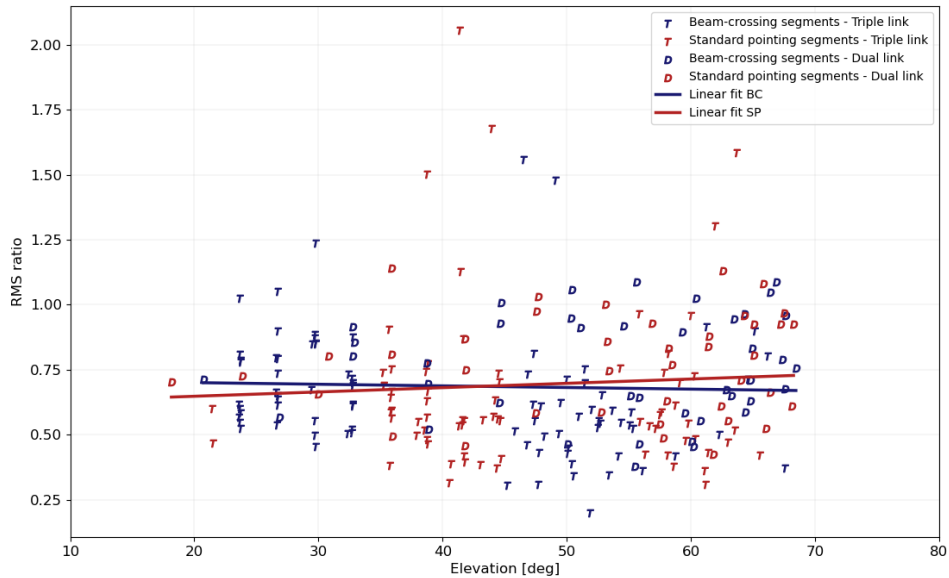

**Figure 8 RMS ratio (TDCS/GNSS) as function of the ground station antenna elevation.**

Figure 7 shows similar and almost constant values of the absolute RMS with respect to the elevation. However, we see from Figure 8 that RMS ratio shows different performances, with the standard pointing performing slightly better at low elevations and the beam-crossing having better performance at higher elevations (> 40°). This could be interesting also in the optics of a possible combined pointing in which the two configurations can be used in the same tracking to optimize the calibration process.

Another relevant correlation was found between the RMS ratio and the observed boundary layer height (BLH) data retrieved from the ECMWF ERA5 database. As discussed before, the boundary layer height plays an important role in implementing the beam-crossing strategy. Specifically, the surface height of the ideal crossing point between the main antenna and the TDCS beams of 700 m was selected as the 70% of the assumed boundary layer height of 1 km[7]. This assumption is based on simulation results which were never tested with real data before.

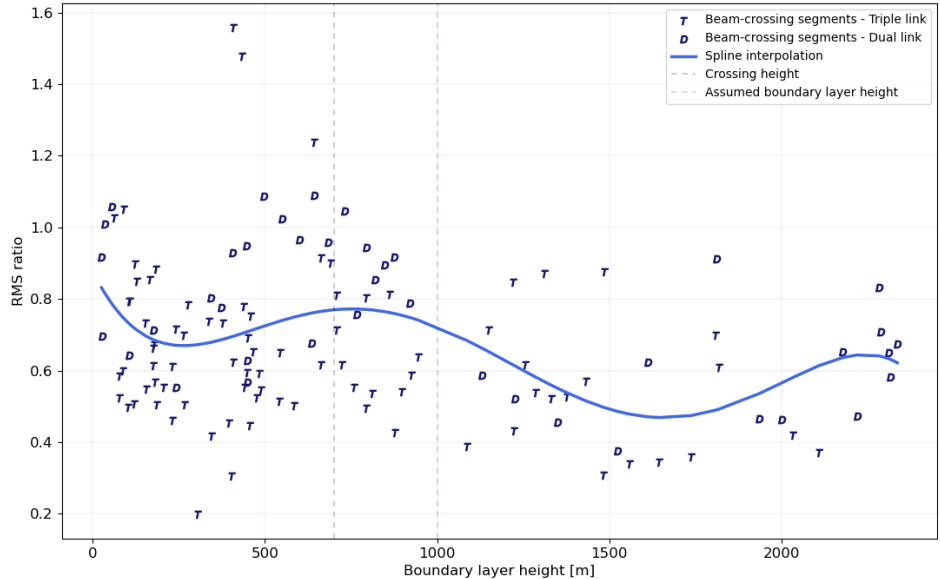

**Figure 9 RMS ratio as a function of the boundary layer height.**

From Figure 9 we first notice that a wide spread of boundary layer height values were encountered during the tracking passes, ranging from a few tenths of meters up to more than 2 km. This consideration already indicates how using a fixed height for all the passes represents a strong assumption. Therefore, future test setups may include variable BLH values for the tracking passes based on statistical analysis of the seasonal and daily BLH fluctuations or base on direct measurements prior to the start of the tracking pass.

Moreover, although the dataset is less extensive at high BLH values, the lowest RMS ratio appears to be reached around 1500 km, corresponding to more than double of the assumed crossing height. This could suggest that the optimal crossing height may be closer to 50% of the instantaneous BLH instead of the assumed value of 70%.

As a final assessment, the beam-crossing campaign results were compared to those obtained during SCE1 and SCE2 when the spacecraft was in superior solar conjunction and the Sun avoidance mode was active, leading to the introduction of an angular offset from the standard pointing. Being the solar conjunction experiments performed in full multifrequency configuration, only the data of test campaign #1 were used for the comparison. Moreover, data affected by adverse weather conditions, were disregarded. For consistency, data collected during SCE1 and SCE2 were also split into 15-minutes segments.

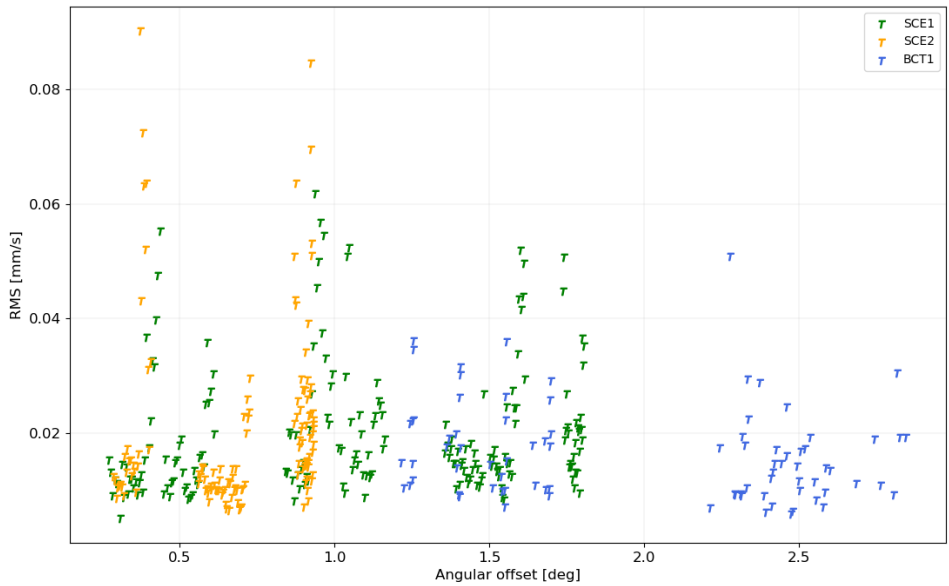

**Figure 10 Absolute RMS residuals as a function of the angular offset from the standard pointing. Blue: beam-crossing; green: SCE1 in Sun avoidance mode; orange: SCE2 in Sun avoidance mode.**

Figure 10 shows the absolute value of the residuals' RMS. We can see that residuals of the SCEs are more spread, reaching values above 0.08 mm/s. Conversely, those of the beam-crossing campaign are all constrained below 0.04 mm/s. Moreover, in the case of the beam-crossing the, RMS slightly decreases as the pointing offset increases.

This feature is also observed in Figure 11, showing the RMS ratio for the three campaigns. From this graph we can see that the RMS ratio of SCE1 data grows with increasing angular offset, reaching values > 1 during periods of low SEP angles. The same trend is not visible for SCE2, which is generally characterized by lower angular offsets and higher SEP angles.

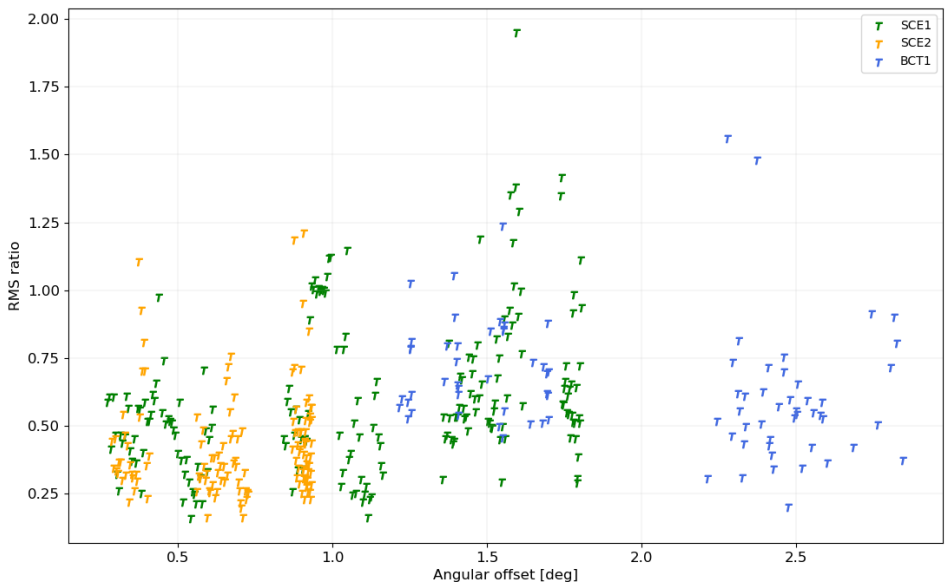

**Figure 11 RMS ratio (TDCS/GNSS) as a function of the angular offset from the standard pointing. Blue: beam-crossing; green: SCE1 in Sun avoidance mode; orange: SCE2 in Sun avoidance mode.**

Figure 11 also shows that the noise reduction of the beam-crossing test increases as the angular offset becomes larger. This could be due to the fact that the smaller angular offsets correspond to observations  performed at low elevation angles, where this technique appears to be less effective as shown in Figure 8.

**CONCLUSIONS**

The beam-crossing is a novel technique which aims at reducing the residual tropospheric Doppler noise when using water vapor radiometers for data calibrations. Described in previous theoretical studies, this technique was tested for the first time in the context of this work using ESA's Tropospheric Delay Calibration System (TDCS) installed at the DS3 complex in Malargue.

A total of 14 tracking passes were analyzed through an orbit determination process using either TDCS-retrieved tropospheric calibrations or GNSS-based calibrations. Specifically, the former were realized by operating the microwave radiometer in two alternative modes, standard pointing and beam-crossing, which were switched at regular time intervals.

A statistical analysis of the Doppler residuals obtained with the two pointing strategies shows that the beam-crossing performance is similar to that of the standard pointing. Modest improvements were observed in the noise reduction and in the Allan deviation values at stability intervals between 100s and 300 s, while slightly degraded performances were observed at low stability intervals. The parameters that were found to affect the results the most were the antenna elevation and the observed boundary layer height during the tracking interval, retrieved from weather prediction databases. These correlations suggest that the simplified assumptions for these first tests (i.e., using a fixed boundary layer height at 700 m from the ground) may need revisiting for the technique to be effective.

Technical challenges related to the complex schedule of BepiColombo operations and to adverse weather conditions limited the number of test tracking passes. Furthermore, the limited use of the Ka-band transponder, prevented the application of the full multi-frequency link for part of the 14 analyzed tracking passes. The illustrated results are thus to be considered extremely preliminary and need to be consolidated by analyzing a much wider dataset that could be obtained by expanding the testing to other missions such as Gaia[25].

The new test data was also compared with the data retrieved during SCE1 and SCE2 passes during superior solar conjunction (SEP < 3°), for which a deliberate pointing offset was introduced to avoid solar intrusion in the TDCS beamwidth (i.e., Sun avoidance mode). For SCE1, we found a positive correlation between the magnitude of the pointing offset and the relative noise reduction introduced by the TDCS calibrations. Conversely, a negative correlation was displayed by the beam-crossing tests. These results suggest that, for comparable angular offsets, the direction in which the offset is applied can have a significant impact on the calibration performances. Future optimizations of the Sun avoidance feature of the TDCS may thus benefit from using a beam-crossing approach. However, further theoretical and test studies are needed to consolidate these findings

In conclusion, the beam-crossing has shown interesting features that opened a path to a possible improvement of the tropospheric calibration system currently used by the ESA DSA-3 antenna in Malargue. Future work should expand the analyzed dataset and feature different approaches for the selection of the crossing height, which could be based on the statistical analysis of past data or real-time predictions of boundary layer height before each tracking pass.

## ACKNOWLEDGMENTS

DB, RLM, MZ and PT are grateful to the Italian Space Agency (ASI) for financial support through the Agreement No. 2022-16 HH.0 for ESA's BepiColombo and NASA's Juno radio science experiments. DB, RLM, MZ, and PT wish to acknowledge *Caltech* and the *NASA Jet Propulsion Laboratory* for granting the University of Bologna a license to an executable version of MONTE Project Edition S/W.